\documentclass[12pt,preprint]{aastex}
\usepackage{epstopdf}
\newcommand{\spit}{\textit{Spitzer Space Telescope}}

\newcommand{\one}{3.6~micron}
\newcommand{\two}{4.5~micron}
\newcommand{\three}{5.8~micron}
\newcommand{\four}{8.0~micron}

\newcommand{\xothreersxot}{1.38}
\newcommand{\xothreeerspxot}{+0.08}
\newcommand{\xothreeersmxot}{-0.08}

\newcommand{\xothreemsxot}{1.213}
\newcommand{\xothreeemspxot}{0.066}

\newcommand{\xothreerpxot}{1.22}
\newcommand{\xothreeerppxot}{+0.07}
\newcommand{\xothreeerpmxot}{-0.07}

\newcommand{\xothreempxot}{11.79}
\newcommand{\xothreeemppxot}{+0.59}
\newcommand{\xothreeempmxot}{-0.59}

\newcommand{\xothreeinclxot}{84.20}
\newcommand{\xothreeeinclpxot}{+0.54}
\newcommand{\xothreeeinclmxot}{-0.54}

\newcommand{\xothreesemixot}{0.0454}
\newcommand{\xothreeesemixot}{0.0008}

\newcommand{\xothreefernhjd}{2,454,449.86816}

\newcommand{\xothreefernper}{3.1915239}

\newcommand{\xothreeccwinn}{0.260}
\newcommand{\xothreeeccwinn}{0.017}
\newcommand{\xothreeccheb}{0.287}
\newcommand{\xothreeeccheb}{0.005}

\newcommand{\xothreecc}{0.277}
\newcommand{\xothreeecc}{0.009}

\newcommand{\xothreeccmarch}{0.278}
\newcommand{\xothreeeccmarch}{0.010}
\newcommand{\xothreeccapril}{0.276}
\newcommand{\xothreeeccapril}{0.009}

\newcommand{\xothreeflone}{0.101\%}
\newcommand{\xothreeeflone}{0.004\%}
\newcommand{\xothreehjdone}{2454908.40094}
\newcommand{\xothreeehjdone}{0.01003}

\newcommand{\xothreephaseone}{0.6720}
\newcommand{\xothreeephaseone}{0.0031}
\newcommand{\xothreegradone}{0.015}
\newcommand{\xothreeegradone}{0.002}

\newcommand{\xothreefltwo}{0.143\%}
\newcommand{\xothreeefltwo}{0.006\%}
\newcommand{\xothreehjdtwo}{2454943.50512}
\newcommand{\xothreeehjdtwo}{0.00608}

\newcommand{\xothreephasetwo}{0.6712}
\newcommand{\xothreeephasetwo}{0.0019}

\newcommand{\xothreeflthr}{0.134\%}
\newcommand{\xothreeeflthr}{0.049\%}
\newcommand{\xothreehjdthr}{2454908.40213}
\newcommand{\xothreeehjdthr}{0.01427}

\newcommand{\xothreephasethr}{0.6724}
\newcommand{\xothreeephasethr}{0.0045}

\newcommand{\xothreeflfour}{0.150\%}
\newcommand{\xothreeeflfour}{0.036\%}
\newcommand{\xothreehjdfour}{2454943.50501}
\newcommand{\xothreeehjdfour}{0.01904}

\newcommand{\xothreephasefour}{0.6712}
\newcommand{\xothreeephasefour}{0.0060}

\newcommand{\xotwogradone}{-0.011}
\newcommand{\xotwoegradone}{0.005}

\newcommand{\onesig}{-2.2}
\newcommand{\twosig}{ 0.3}
\newcommand{\threesig}{-0.8}
\newcommand{\foursig}{-1.7}

\newcommand{\xothrinso}{2.01}

\renewenvironment{thebibliography}[1]{%
\begin{oldthebibliography}{#1}%
\setlength{\parskip}{0ex}%
\setlength{\itemsep}{0ex}%
}%
{%
\end{oldthebibliography}%
}

\shorttitle{\objectname[NAME XO-3]{XO-3}Thermal emission}
\shortauthors{Machalek et al.}

\begin{document}

\title{Thermal Emission and Tidal Heating of the Heavy and Eccentric Planet XO-3b}

\author{
Pavel~Machalek\altaffilmark{1,2},
Tom~Greene\altaffilmark{1},
Peter~R.~McCullough\altaffilmark{3},
Adam~Burrows\altaffilmark{4},
Christopher~J.~Burke\altaffilmark{5},
Joseph~L.~Hora\altaffilmark{5},
Christopher~M.~Johns-Krull\altaffilmark{6},
Drake~L.~Deming\altaffilmark{7}
}

\email{pavel.machalek@nasa.gov}
\altaffiltext{1}{NASA Ames Research Center, MS 245-6, Moffett Field, CA 94035}
\altaffiltext{2}{Bay Area Environmental Research Institute, 560 Third St West, Sonoma, CA 95476}
\altaffiltext{3}{Space Telescope Science Institute, 3700 San Martin Dr., Baltimore MD 21218}
\altaffiltext{4}{Department of Astrophysical Sciences, Princeton University, Princeton, NJ 08544}
\altaffiltext{5}{Harvard-Smithsonian Center for Astrophysics, 60 Garden St., Cambridge, MA 02138} 
\altaffiltext{6}{Deptartment of Physics and Astronomy, Rice University, 6100 Main Street, MS-108, Houston, TX 77005}
\altaffiltext{7}{Planetary Systems Laboratory, NASA/GSFC, Code 693.0, Greenbelt, MD 20771} 

\begin{abstract}
We determined the flux ratios of the heavy and eccentric planet XO-3b to its parent star in the four IRAC bands of the Spitzer Space Telescope: \xothreeflone~$\pm$ \xothreeeflone~at \one; \xothreefltwo~$\pm$ \xothreeefltwo~at \two; \xothreeflthr~$\pm$ \xothreeeflthr~at \three~and \xothreeflfour~$\pm$ \xothreeeflfour~at \four. The flux ratios are within [\onesig,\twosig, \threesig, \foursig]-$\sigma$ of the model of XO-3b with a thermally inverted stratosphere in the \one,~\two,~\three~and \four~channels, respectively.  XO-3b has a high illumination from its parent star ($F_p \sim $(1.9 - 4.2) $\times$ 10$^{9}$ ergs cm$^{-2}$ s$^{-1}$) and is thus expected to have a thermal inversion, which we indeed observe. When combined with existing data for other planets, the correlation between the presence of an atmospheric temperature inversion and the substellar flux is insufficient to explain why some high insolation planets like TrES-3 do not have stratospheric inversions and some low insolation planets like XO-1b do have inversions. Secondary factors such as sulfur chemistry, atmospheric metallicity, amounts of macroscopic mixing in the stratosphere or even dynamical weather effects likely play a role. Using the secondary eclipse timing centroids we determined the orbital eccentricity of XO-3b as e = $\xothreecc \pm \xothreeecc$. The model radius-age trajectories for XO-3b imply that at least some amount of tidal-heating is required to inflate the radius of XO-3b, and the tidal heating parameter of the planet is constrained to Q$_p \lesssim$ 10$^6$. 

\end{abstract}

\keywords{stars:individual(XO-3) --- binaries:eclipsing --- infrared:stars --- planetary systems}


\slugcomment{Accepted for publication in The Astrophysical Journal}

\section{Introduction}
\label{xo3_intro}

The study of hot Jupiter atmospheres is maturing. In particular, low resolution spectra and broadband spectral energy
 distributions have been assembled from high precision photometry of
 Hot-Jupiter's day and night sides using the Spitzer Space Telescope's
 InfraRed Array Camera (IRAC) \citep[][]{knutson07b, knutson_tres4,
 tin07,charb08,machalek08, machalek09, odonovan09,desert09,
 todorov09,fressin09, christi09}, Infrared Spectrograph (IRS)
 \citep{grill09} and Multi Band Imaging Spectrometer (MIPS)
 \citep{knutson09_phase} as well the Hubble Space Telescope
 \citep{swain08,swain08b,swain09}.\\

Upper atmospheres of hot Jupiters are currently thought to be split
 into two classes depending on the stellar insolation at their
 substellar points: planets with substellar flux higher than F$_p
 \gtrsim 10^{9}$ erg cm$^{-2}$ s$^{-1}$ should posses temperature
 inversions in their stratosphere as the intense stellar radiation is
 absorbed by upper atmospheric gaseous absorbing species
 \citep{hubeny03,burr07b,fort06,fort07b,spiegel09}.  Planets with
 insolation fluxes F$_p$ $\sim$ 0.5-1.0 $\times$ 10$^{9}$ ergs cm$^{-2}$
 s$^{-1}$ like XO-2b, HAT-P-1, OGLE-TR-113, and WASP-2 are in a
 transition zone between atmospheres with or without a stratosphere.
 Secondary effects like sulfur chemistry and atmospheric metallicity
 \citep{zahn09}, amounts of macroscopic mixing in the stratosphere
 \citep{spiegel09} or even dynamical weather effects
 \citep{showman09,rauscher09} could determine the stratospheric
 temperature profiles of these transition planets.

 \begin{figure}[!th]
  \centering
  \includegraphics[width=0.6\textwidth]{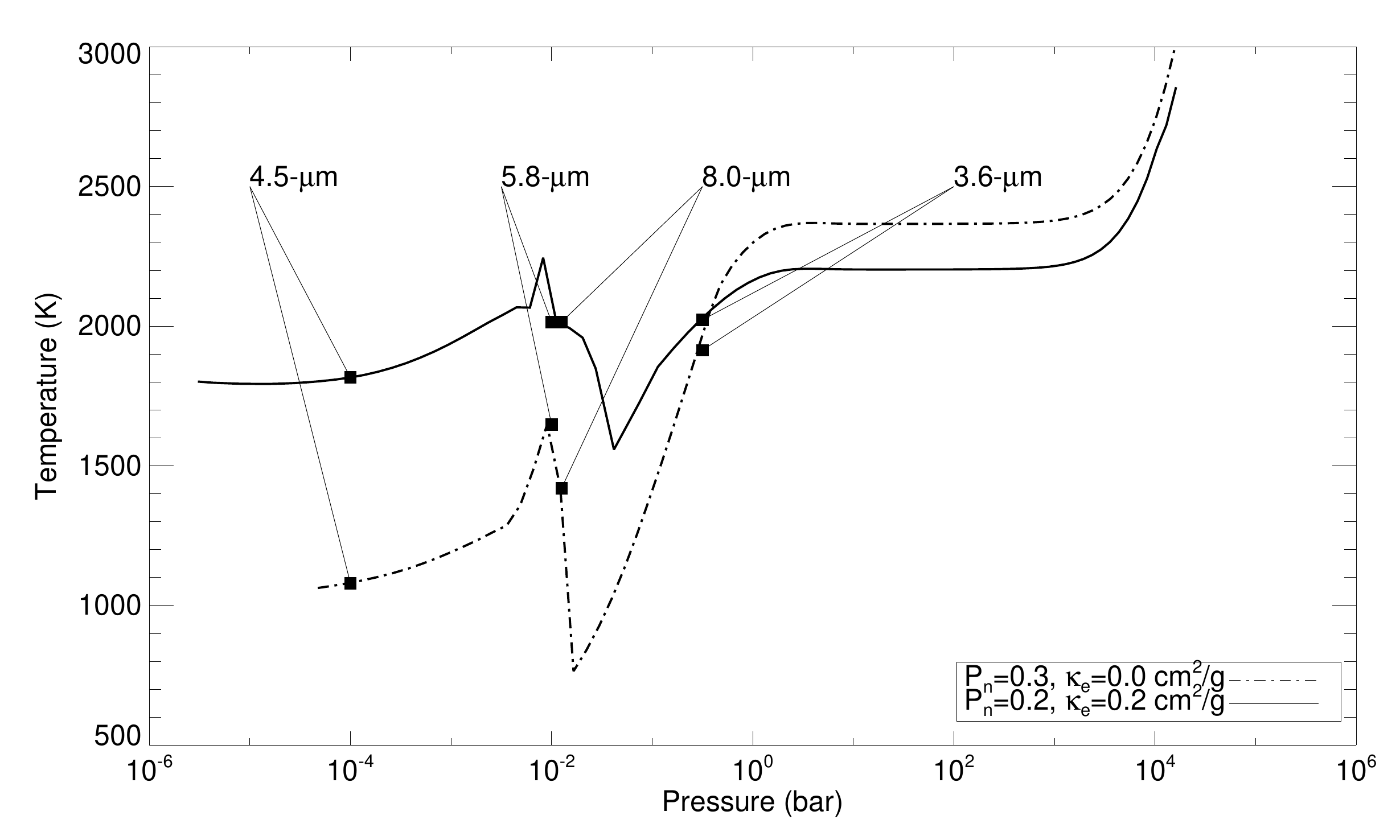}
  \caption[Temperature - Pressure profiles for XO-3b]{(top) Temperature / Pressure profiles for the atmosphere of XO-3b following the methodology of \citet{burr07,burr07b,spiegel09} for heat redistribution parameter P$_{n}$=0.3 with no upper atmospheric optical absorber (dot-dashed line) and a corresponding model with a uniform upper atmospheric absorber (solid line) (absorption coefficient $\kappa_{e}$=0.2 cm$^{2}$/g and heat re-distribution parameter P$_{n}$=0.2) with depths corresponding to emission in the IRAC channels denoted. Both Temperature/Pressure profiles are calculated for XO-3b orbital distance $a$ = \xothreesemixot~AU \citep{winn08} and stellar insolation $F_p \sim $\xothrinso $\times$ 10$^{9}$ ergs cm$^{-2}$ s$^{-1}$. See \S \ref{anlxo3} for more details.  \label{xo3b_tp} }
\end{figure}

XO-3b is a hot Jupiter with a high mass M$_{p}$ = 11.79 $\pm$ 0.59 M$_{Jup}$~\citep{winn08,xo3}, which is close to the deuterium burning limit and has one of the highest observed surface gravities, g = 209 m s$^{-2}$ amongst the known transiting planets. Its \xothreefernper~day long orbit around the parent star XO-3
 (spectral type F5V, d = 260 $\pm$ 23 pc, \citet{xo3} ) has significant
 eccentricity e = \xothreeccheb $\pm$ \xothreeeccheb~\citep{heb08}),
 which causes stellar irradiance to vary three-fold over the entire
 orbit and causes the secondary eclipse to shift in time from
 half-phase.

Furthermore \citet{liu08} estimated the amount of tidal energy dissipation rate contributing to the inflated radius of XO-3b \citep[R$_p$ = 1.217 $\pm$ 0.073 R$_{Jup}$;][]{winn08} assuming the age of XO-3b t= 2.82 $^{+0.58}_{-0.82}$ Gyr \citep{winn08}. \citet{liu08} concluded that the radius-age relationship for XO-3b is consistent to within 1.0-$\sigma$ with no internal heat source (i.e. no tidal heating) or tidal heating dissipation with dimensionless tidal heating parameter Q$_p \gtrsim$ 10$^6$ as defined by \citet{gold66}. By determining the exact timing of the secondary eclipse in our 4 infrared light curves obtained with \spit~IRAC we will refine the orbital eccentricity of XO-3b and constrain the amount of tidal heating (if any) responsible for inflating the planetary radius. 
In addittion to its high mass and significant orbital eccentricity, XO-3b was also the first planet with detected and confirmed non-zero sky projection angle $\lambda$ = 37.3 $\pm$ 3.7 deg between the orbital axis and stellar rotation axis obtained from the Rositter-McLaughlin effect \citep{heb08,winn09}, currently thought to be a result of planet-planet scattering \citep{nagasawa08,juric08}.\\

The substellar point flux at XO-3b is $F_p \sim $(1.9 - 4.2) $\times$ 10$^{9}$ ergs cm$^{-2}$ s$^{-1}$. The exact value depends on the adopted stellar and planetary mass and radius, which are still uncertain \citet{liu08}, as well as the changing distance distance from the star due to an eccentric orbit. However this range of substellar point flux is clearly consistent with a prominent thermal inversion in the stratosphere. Figure \ref{xo3b_tp} shows Temperature- Pressure models of \citet{burr07,burr07b,spiegel09} and the predicted thermal inversion in the stratosphere and the negative temperature gradient in the upper atmosphere of XO-3b. \\

By obtaining the light curve of XO-3b in the 4 IRAC (4-8 microns) channels on the \spit~and determining the depth and timing of the secondary eclipse in multiple wavelengths, we will be able to constrain the upper atmospheric temperature structure of XO-3b, refine the orbital eccentricity of the planet from the secondary eclipse timing centroids and hence its tidal heating rate, which could be responsible for inflating the radius of XO-3b. 
 The Cold Spitzer IRAC observations in this work  will provide a firm observational and theoretical foothold on the properties of the XO-3b atmosphere during the secondary eclipse and serve as comparison for future full orbit observation of XO-3b with Warm Spitzer similar to previous extended duration phase curves of hot Jupiters \citep{knutson07,knutson09_phase,knutson09_hd149026,laughlin09}. 
Since there is a strong water band near IRAC 5.8 micron, coverage in all four IRAC bands will test for transitions between water in emission and in absorption, which can not be observed with Warm Spitzer. Furthermore as Fig. \ref{xo3b_tp} illustrates, there is a steep temperature gradient between depths corresponding to emission in IRAC 5.8/8.0 micron channels, which can be uniquely studied with Cold Spitzer or otherwise with JWST in the future.   The 5.8 and 8.0 micron channel planet/star flux ratios will further be correlated with the  3.6/ 4.5 micron flux ratio to test the two signatures of stratospheres.

\section{Observations \& Data Analysis}
The InfraRed Array Camera \citep[IRAC;][]{fazio04} has a field of view of
5.2$\arcmin$ $\times$ 5.2$\arcmin$ in each of its four bands. Two adjacent
fields are imaged in pairs (3.6 and 5.8 microns; 4.5 and 8.0 microns). The
detector arrays each measure 256 $\times$ 256 pixels, with a pixel size of
approximately 1.22$\arcsec$ $\times$ 1.22$\arcsec$. We closely repeat the data analysis of \citet{machalek08,machalek09} with modifications and improvements mentioned in the text. \\

We have observed \objectname[TYC 3727-1064-1]{XO-3}~ system in all 4 channels in two separate Astronomical
Observing Requests (AORs) in two different sessions: the 3.6 and 5.8 micron
channels for 6.9 hours (with 2.9 hour long secondary eclipse) on UT 2009 March 17 (AOR 31618560) and the 4.5 and
8.0 micron channels for 6.9 hours on UT 2009 April 21 (AOR 31618816) with a 30-minute preflash on a bright uniform part of \objectname{NGC1569}. We used the full array 2s+2s/12s frame time in the stellar mode in which the \one~and \two~bands are exposed for two consecutive 2s exposures while the \three~and \four~bands are integrating for 12s to prevent detector saturation.


The \two~and \four~ time series has been preflashed with a bright uniform extended target to prevent the initial ``ramp-up'' effect \citep{charb05,dem05,knutson07b,machalek08,machalek09}, consequently no data points were removed from the beginning of the time series. The \one~and \three~time series however were obtained with no pre-flashing and hence exhibit an initial charge build up which is consistently removed during our detector effect removal.

\subsection{InSb Detectors}
\label{xo3_insb}
We have repeated our methodology from \citet{machalek09} by performing aperture photometry on the \one~and \two~time series with radii between 2.5 and 6.0 pixels. In order to test whether our secondary eclipse depths and centroid timings depend on aperture radius, we have repeated the entire data reduction for aperture radii between 2.5 and 6.0 pixels in 0.5 pixel increments and obtained consistent results for different apertures. We have improved our photometry pipeline by obtaining the stellar centroids from flux-weighted position of a 5 $\times$ 5 pixel square centered on the peak stellar pixel (method suggested by Sean Carrey, private communication). Since our starting point was the BCD images produced by the pipeline version 18.7, cosmic rays were already rejected. The heliocentric modified Julian date at Spitzer spacecraft position recorded in the header keyword ``HMJD\_OBS'' did not necessitate our previous calculations of spacecraft positions \citep{machalek08,machalek09}. 
 
We have chosen the aperture radii based on the RMS of residuals after detector effects and the secondary eclipse were removed. We used an aperture of radius 3.0 pixels for the \one~time series of XO-3, which had an RMS 0.0034 for out of transit points after decorelation. This is essentially Poisson noise limited, being only 1.01 higher than the predicted noise based on source brightness, detector read noise and gain. Similarly the \two~time series of XO-3 was obtained from 3.0 pixel radius aperture photometry which had the lowest RMS of 0.0049 which is 1.08 times higher than the predicted noise. The appropriate aperture corrections were applied to the photometry as specified by the Spitzer Data Handbook. 

As is evident from Fig. \ref{fig:instruxo3}, the \one~time series exhibits a prominent flux variation with magnitude of $\sim$ 0.8 \%, which is a well studied instrumental effect  \citep{charb05,morales06,machalek08,knutson_tres4,machalek09,desert09} due to sub-pixel sensitivity variations caused by spacecraft position drift of 0.1 - 0.3 arcsec over a period of $\sim$ 3000 seconds, which makes the star move on the pixel. The \two~time series, however, has negligible flux variations, probably due to a chance positioning on a pixel phase with a flat response curve (pixel reference: 126.46; 128.78). This pixel could be useful in planning for extended duration observations with Warm Spitzer. \citet{desert09} has noted a similar pixel with a flat response function at pixel coordinates [147.20; 198.25].    

Our removal of the systematic effects and eclipse curve fitting closely follows the methodology of \citet{machalek09}. The subpixel intensity variations in the \one~and \two~ time series are detrended as a linear function of subpixel positions of the stellar centroid  $x$,~$y$,~$x^2$,~$y^2$, a linear function of time t, plus a constant for each of the two InSb channels:

\begin{equation}
  \label{eq:subpixel36}I_{3.6 micron}=~ 1.0 +~b_{1}x +~b_{2}y~+~b_{3}x^2 +~b_{4}y^2~+~b_{5}t, 
\end{equation}    

\begin{equation}
  \label{eq:subpixel45}I_{4.5 micron}=~ 1.0 +~b_{1}x +~b_{2}y,
\end{equation} 

\begin{figure}[!tph]
\centering
\includegraphics[totalheight=6cm]{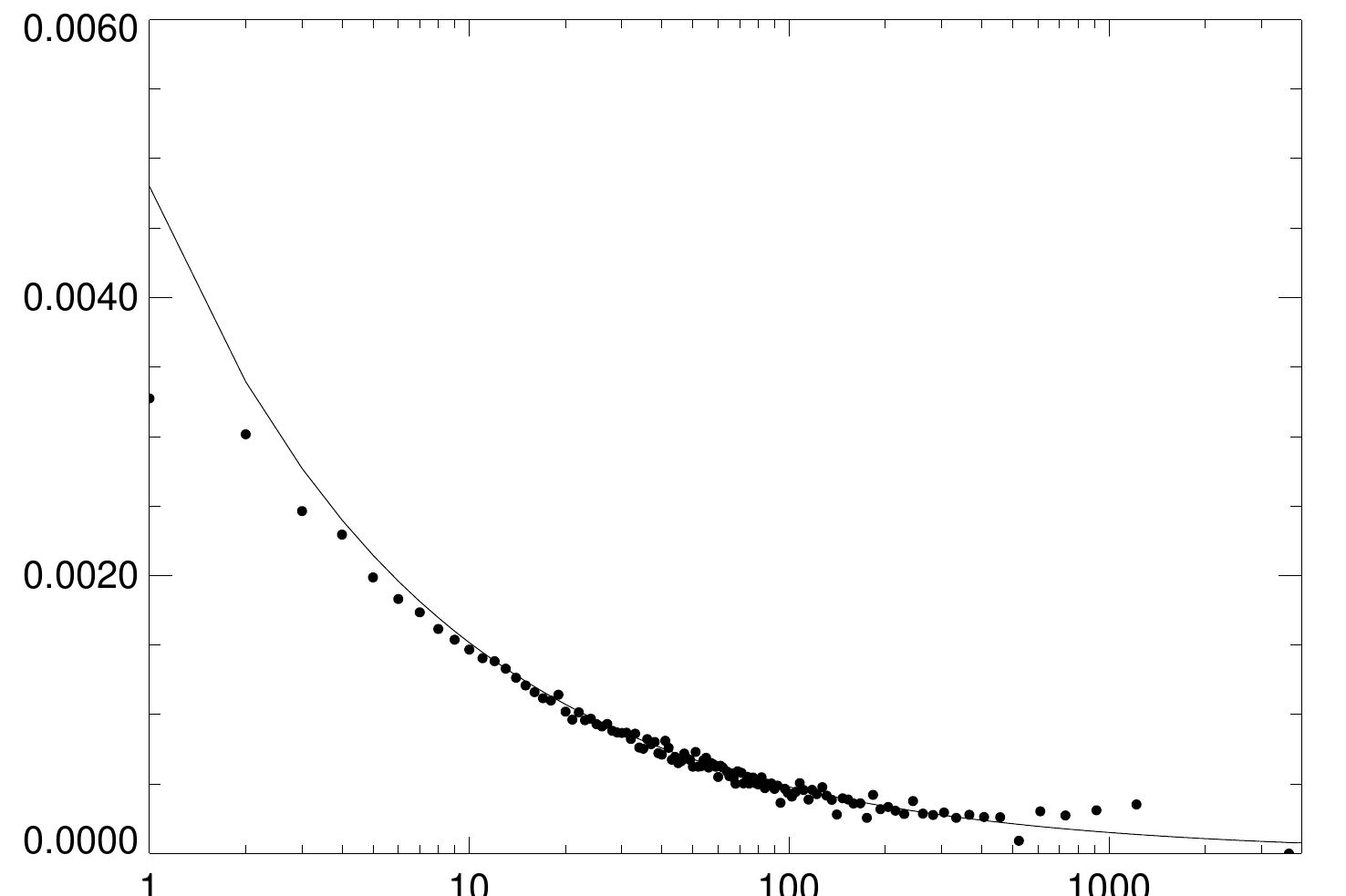}
\caption[XO-3 4.5 micron light curve bins vs RMS]{Root mean square of binned light curve points after detector removal and secondary eclipse fitting for the 4.5 micron photometry of XO-3 vs. the number of points per bin. The sub-pixel phase sensitivity variations were removed using Eq. \ref{eq:subpixel45}. The solid line is proportional to the number of residual points per bin N$^{-1/2}$.}
\label{fig:xo3_resbin}
\end{figure}

We tried adding higher order terms of $x$ and $y$, a cross terms of  $x \times y$, and a linear term linear in t to the \two~ time series decorelation. However, adding terms did not decrease the $\chi^{2}$ or change the secondary eclipse depth or centroid timing in the \two~ time series, so we chose only two degrees of freedom (Eq. \ref{eq:subpixel45}) for the \two~ time series decorelation. 
Furthermore as can be seen from Fig. \ref{fig:xo3_resbin}, the binned residuals in the decorrelated and fitted 4.5 micron light curve of XO-3 scale as N$^{-1/2}$, where N is the number of points per bin. Since the binning of the residuals scales as  N$^{-1/2}$ we can conclude that negligible systamtic errors remain in the decorelated light curve.

We fit the secondary eclipse with the formalism of  \citet{agol02} with no stellar limb darkening  and adopt the stellar and planetary parameters of \citet{winn08}:

R$_{\star}$ = \xothreersxot $^{\xothreeerspxot}_{\xothreeersmxot}$ R$_{\sun}$, M$_{p}$ = \xothreempxot $^{\xothreeemppxot}_{\xothreeempmxot}$ M$_{Jup}$, R$_{p}$ = \xothreerpxot $^{\xothreeerppxot}_{\xothreeerpmxot}$ R$_{Jup}$\footnote{1 R$_{Jup}$ = 71,492 km.}, $i$ =~\xothreeinclxot$^{\xothreeeinclpxot}_{\xothreeeinclmxot}$ degrees, and $a$ = \xothreesemixot~$\pm$~\xothreeesemixot~AU with ephemeris:

\begin{equation}
  \label{eq:ephmxo3} T_{c}(E) = \xothreefernhjd (HJD) + E(\xothreefernper~days) \, .
\end{equation}

We fit the 5 baseline parameters of Eq. \ref{eq:subpixel36} and the 2 baseline fitting parameters Eq.~\ref{eq:subpixel45} concurrently with the secondary eclipse depth $\Delta F$ and the phase of the eclipse centroid $\Phi$ for a total of 7 and 4 fitting parameters, respectively. This was done to properly account for the way in which systematic effects removal affects the secondary eclipse fitting. The best parameter solutions were obtained by using a Monte Carlo Markov Chain (MCMC) with 10$^{5}$ iterations  \citep{gregory05,mark09} with ratio of jumps between 20-40 \%. 
The best fit parameter values were obtained by discarding the first 20\% of the iterations to prevent initial conditions from influencing the results and adopting the median of the distribution of each parameter as the best fit value. These values are reported in Table \ref{tbl1xo3} with errors obtained from symmetric 66.8\% contours around the median of the posterior probability distribution of the MCMC runs. The decorelated best-fit light curves are depicted in Fig. \ref{fig:fitxo3} binned in 3.5 minute intervals. Note, however, that all our analysis is performed on the unbinned data.  \\

We find that the XO-3 \one~ time series shows a linear flux increase with a slope of  b$_{4}$ = \xothreegradone \% $\pm$ \xothreeegradone \% per hour which is consistently removed from our photometry, but inconsistent with the slope of XO-2 at \one~ of  b$_{4}$ = \xotwogradone \% $\pm$ \xotwoegradone \% per hour \citep{machalek09}. This flux decrease has been attributed by \citet{machalek09} and \citet{knutson_tres4} to an instrumental effect on the In:Sb detectors. When we added a linear time term $b_{3}t$ to the decorelation of the \two~ time series of XO-3 in Eq.~\ref{eq:subpixel45}, its value was consistent with zero. Thus we omitted a linear time term $b_{3}t$ from the final analysis. 

\subsection{Si:As Detectors}
\label{siasxo3}
The \three~and \four~ time series is recorded with Si:As detectors, which have a  different set of systematic effects from the \one~and \two~ InSb detectors. We have performed aperture photometry on the \three~and \four~ images with aperture radii ranging from 3.0 to 6.0 pixels, choosing the aperture radius with the lowest RMS of the residuals after systematic effects and the secondary eclipse were removed. This resulted in an aperture of radius 3.5 pixels for the \three~time series with a detrended RMS of 0.0055 (42\% higher than Poisson noise) and an aperture radius of 4.5 pixels for the \four~time series with a detrended RMS of 0.0049 (60\% higher than Poisson noise). No points were removed from the beginning of either the \three~or \four~ time series.    \\

\begin{figure}
\centering
\includegraphics[totalheight=0.9\textwidth]{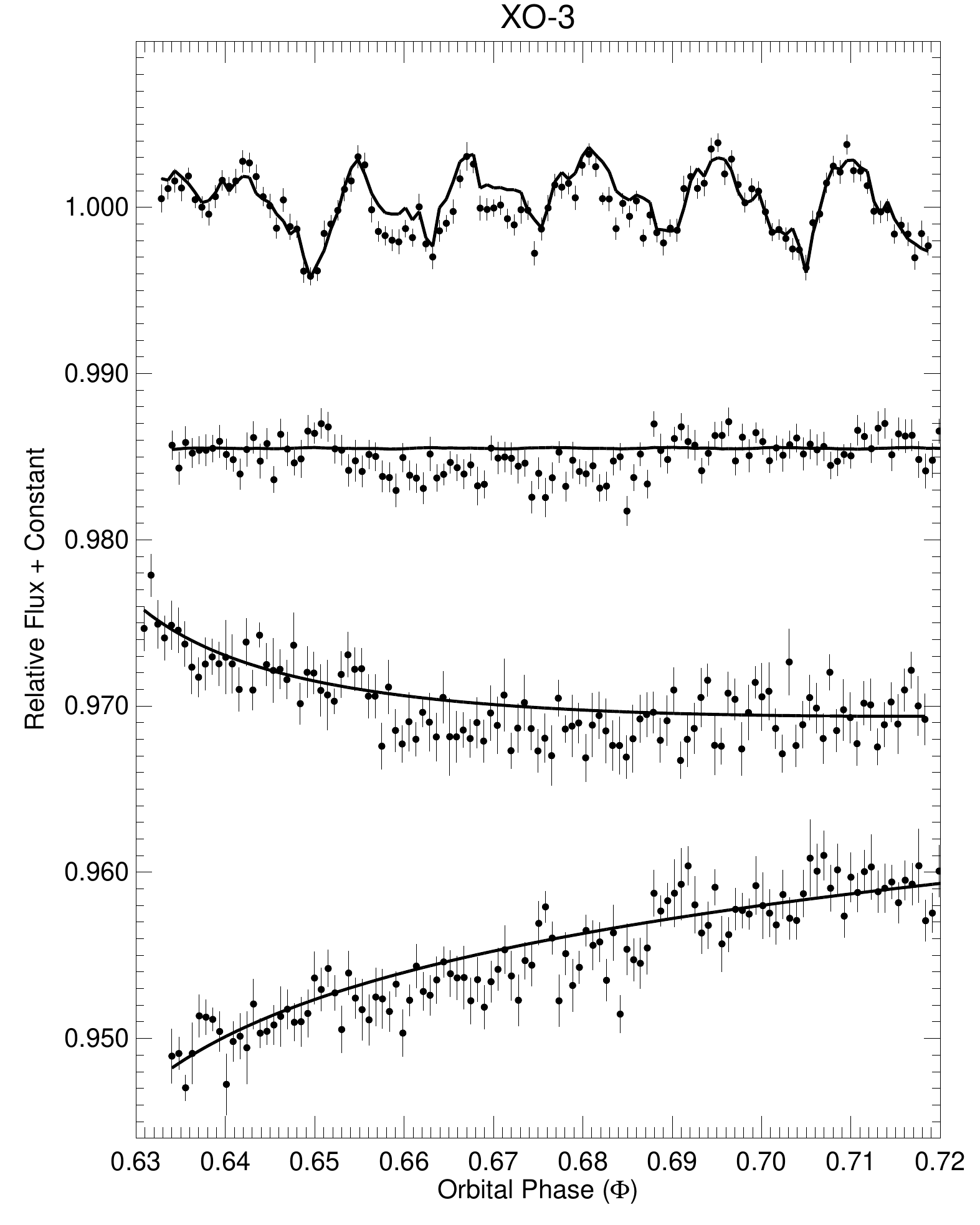}
\caption[Raw IRAC XO-3 time series]{\small (left) Secondary eclipse observations of \protect\objectname{XO-3}~ with 
IRAC on \spit~ in \one, \two, 
\three~and \four~channels (from top to bottom) binned in 3.5-minute intervals 
and normalized to 1 and offset for clarity. Note, however, that all our analysis is performed on the unbinned data. The overplotted solid  lines show the corrections for the detector effects (see text).}
\label{fig:instruxo3}
\end{figure}

A well studied instrumental effect of the Si:As arrays is the gain variations of individual pixels over time, which result in flux decrease/increase in the light curve \citep[e.g.][]{dem05,knutson07,knutson07b, machalek08,desert09}, quite unlike the pixel position dependent flux effect in the InSb \one~and \two~arrays. \citet{machalek09} and \citet{laughlin09} have reported that the gain variations in the \three~and \four~ channels and resultant flux trends in the light curves differ for the two components of a binary star, which have the same brightness and similar colors, suggesting that relative placement of the stellar centroid with respect to the edges of the pixels determines the Si:As detector pixel response. The gain variations can be clearly seen in the \three~and \four~ light series in Fig. \ref{fig:instruxo3}: a nonlinear decrease in brightness in the \three~ light series and a nonlinear flux increase in the \four~time series. \\

\begin{figure}
\centering

\includegraphics[totalheight=0.9\textwidth]{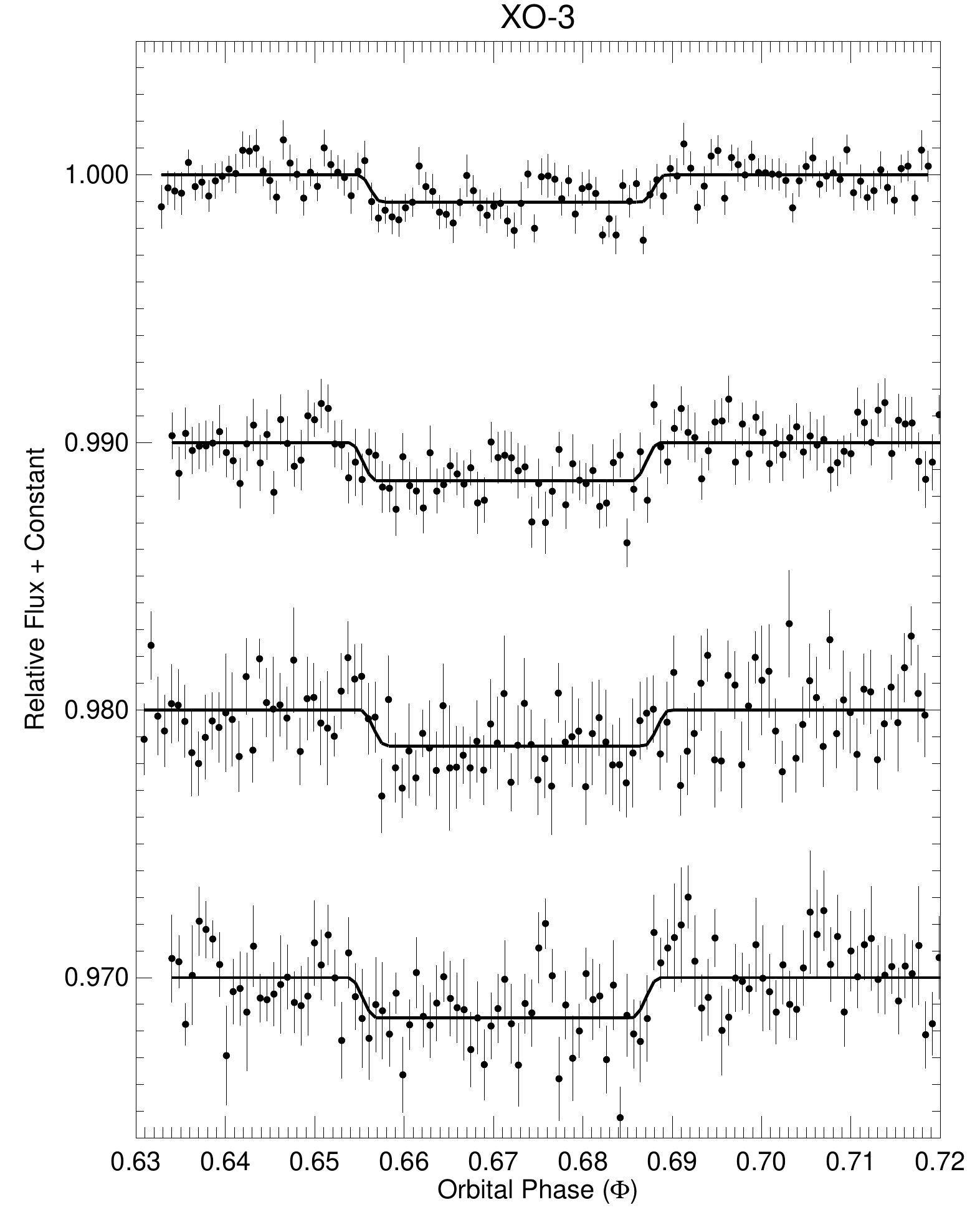}

\caption[Corrected IRAC XO-3 photometry]{Secondary eclipse of \protect\objectname{XO-3b} around the star XO-3 observed with IRAC on \spit~in 3.6, 4.5, 5.8, and 8.0 micron channels (top to bottom) corrected for detector effects, normalized and binned in 3.5-minute intervals and offset for clarity. The best-fit eclipse curves are overplotted. Note, however, that all our analysis is performed on the unbinned data.}
\label{fig:fitxo3}
\end{figure}

To remove the nonlinear flux variation inherent to the Si:As detector, we fit the secondary eclipse depth $\Delta F$ along with the eclipse centroid phase $\Phi$ concurrently with the 3 ``ramp'' decorelation coefficients as follows:
 
\begin{equation}
  \label{eq:rampxo3}I_{model} = a_{1}  + a_{2} \times~  ln(\Delta t +0.05) + a_{3} \times~  ln(\Delta t +0.05)^2 
\end{equation}
where $I_{model}$ is the normalized model flux and $\Delta t$ is the time in days since the beginning of the integration (constant of +0.05 inserted to avoid singularity at $\Delta t$=0.). 
We fit the 5 parameters (2 for the eclipse and 3 for the ``ramp'' in Eq. \ref{eq:rampxo3}) for the \three~and \four~time series concurrently using 10$^{5}$ MCMC runs with errors adopted as the 66.8 \% contours around the median of the posterior distribution of the MCMC runs for each parameter. 
To ensure that our results are not dependent on the aperture radius we have repeated the MCMC runs for all aperture radii between 3.0 and 6.0 piels in 0.5 pixel increments and found the timing centroids to be consistent. The secondary eclipse depths were, however, found to vary by about 1-$\sigma$ for photometry with aperture radii between 3.0 and 6.0 pixels. 
Hence to be conservative, as stated above, we have adopted the secondary eclipse depths from the aperture photometry with the lowest RMS of residuals after eclipse removal.  
These aperture radii were 3.5 pixels for the \three~time series and 4.5 pixels for the \four~time series. We adopted uncertainties as the upper and lower envelope of the eclipse depths with their uncertainties for photometry with aperture radii between 3.0 and 6.0 pixels. Note, however, that these large, conservative uncertainties of the \three~and \four~ eclipse depths (  $\Delta F_{\three}$= \xothreeflthr~$\pm$ \xothreeeflthr~and  $\Delta F_{\four}$= \xothreeflfour~$\pm$ \xothreeeflfour ) still allow us to distinguish between the two models for the upper atmospheric temperature structure of XO-3b (see Fig. \ref{fig:atmoxo3}). 
The final results are reported in Table \ref{tbl1xo3}, and the decorelated time series was binned in 3.5-minute bins for viewing clarity in Fig. \ref{fig:fitxo3}.  


\section{Discussion}

The IR light curves presented in this work allow for the determination of the exact timing of the secondary eclipse centroid such that the orbital eccentricity can be refined more accurately than from the radial velocity (RV) curve. The temperature structure of the upper atmosphere of XO-3b can also be determined from the light curves by comparing the secondary eclipse depths (i.e. the planet/star contrastratios) to atmospheric models. A refined eccentricity determines the rate of tidal heating of the planet and can help explain the inflated radius of XO-3b.  
 
\subsection{Tidal heating rate and the radius of the planet}
\label{xo3_ecc}

A subset of transiting extrasolar giant planets (EGPs) have radii larger than standard models can accommodate \citep{guillot96,bodenheimer01,bodenheimer03,chabrier04,ibgui09}. Numerous explanations have been suggested as sources of the inflated radii of EGPs (see \citet{fort09} for a review).  Working in opposition to any inflation mechanism, heavy-element inner cores lead to smaller planetary radii compared to pure H/He objects \citep{burr07c,fort07,baraffe08}. Tidal inflation has been a popular explanation, as the dissipation of orbital energy into the inner regions of a planet can lead to inflated radii \citep{bodenheimer01,bodenheimer03,liu08}. Radius-age trajectories for extra-solar giant planets (EGPs) are presented by \citet{liu08} to explain the inflated radius of several planets, including XO-3b, which is larger than theoretical predictions.\\

\citet{liu08} have investigated the radius of XO-3b as a function of planetary age t, planetary radius R$_{p}$, planetary mass M$_{p}$, orbital eccentricity e, planetary metallicity [Fe/H], and tidal heating parameter Q$_{p}$. 
They conclude that for the parameters adopted from the photometric follow-up of XO-3b by \citet{winn08}, which are used in our study (see \S \ref{xo3_insb}), the radius R$_{p}$ = \xothreerpxot $^{\xothreeerppxot}_{\xothreeerpmxot}$ R$_{Jup}$ adopted by \citet{winn08} is consistent within 1-sigma to either no internal heat source or tidal energy dissipation with tidal heating parameter Q$_{p}$ $\gtrsim$ 10$^{6.0}$. This is the heating parameter for the adopted eccentricity based upon RV measurements from \citet{winn08}, e = \xothreeccwinn $\pm$ \xothreeeccwinn. We have also adopted these parameters of \citet{winn08} up to this point in this study (see \S \ref{xo3_insb}).\\

We refine the eccentricity e of XO-3b using the weighted average of the secondary eclipse timing centroids from Table \ref{tbl1xo3} using the displacement from half orbital phase as a measurement of eccentricity (e.g. \citet{kopal59} Eq. 9.23):

\begin{equation}
  \label{eq:eccxo3}  2\pi \Phi \simeq \pi + 2e\times cos(\omega)(1+csc^{2}(i))+ ...
\end{equation}

where $e$ is the eccentricity, $\Phi$ is the orbital phase of the time centroid of the secondary eclipse, $\omega$ is the longitude of periastron, and i is the planetary orbit inclination. 
Using argument of pericenter $\omega$=345.8 deg $\pm$~7.3 deg and inclination i = 84.20 deg $\pm$~0.54 deg from \citet{winn08}, we derive a refined eccentricity of the XO-3b system from our secondary eclipse timings:

\begin{equation}
  \label{eq:eccxo3_new}  e = \xothreecc \pm \xothreeecc 
\end{equation} 

with uncertainties formally propagated through Eq. \ref{eq:eccxo3}. Taken individually the March 2009 secondary eclipse phase centroids from Table \ref{tbl1xo3} imply an eccentricity of $\xothreeccmarch \pm \xothreeeccmarch$ and the April 2009 secondary eclipse phase centroids imply eccentricity of  $\xothreeccapril \pm \xothreeeccapril$, which are consistent with each other. The intriguing possibility of eccentricity changing on a timescale of months will be further studied during the Warm Spitzer mission phase observations of XO-3b in spring 2010, when both the transit and secondary eclipse will be observed. \\   

The refined value of XO-3b eccentricity e = \xothreecc $\pm$ \xothreeecc~is 1.0-$\sigma$ higher than the \citet{winn08,xo3} eccentricity e = \xothreeccwinn~$\pm$ \xothreeeccwinn. It is also 2.0-$\sigma$ lower than the radial velocity derived eccentricity e = \xothreeccheb $\pm$ \xothreeeccheb, \citet{heb08}. The tidal heating of XO-3b, which is a strong function of eccentricity, can inflate the radius of the planet.  To estimate the relevance of tidal heating to the energy budget of XO-3b, we evaluate the ratio of the tidal energy dissipation rate to the insolation rate from the parent star \citep{liu08}:

\begin{equation}
\begin{array}{ll}
  \label{eq:eccxo3_ratio}  \frac{\dot{E}_{tide}}{\dot{E}_{insolation}} = \frac{GM_{\star} \mu f(e) }{\pi F_p R_p^2 a \tau_{circ}} &\\
\sim 6.9 \times 10^{-5} (\frac{e}{0.01})^2  [\frac{f(e)}{e^2}]  (\frac{Q_p}{10^5})^{-1}  (\frac{M_{\star}}{M_{\sun}})^{5/2} (\frac{R_p}{R_j})^3 (\frac{a}{0.05 AU})^{-15/2}  (\frac{F_p}{10^9~ergs~cm^{-2}~s^{-1} })^{-1}
\end{array}
\end{equation}

where $\dot{E}_{tide}$ is the tidal energy dissipation within the planet's rest frame, $\dot{E}_{insolation}$ is the insolation rate of the planet $\dot{E}_{insolation}$ = $\pi R_p^2 F_p$, where $R_p$ is the radius of the planet; $F_p$ is the stellar flux at the planet's substellar point;  $\mu$ is the reduced mass $\mu \equiv \frac{M_{\star}M_{planet}}{M_{\star}+M_{planet}}$; $\tau$ is the circularization timescale as defined by \citet{liu08}; 
e is orbital eccentricity; f(e) is a function of eccentricity f(e) $\equiv \frac{2}{7}(h_3(e) - 2h_4(e) +h_5(e))$. The terms of f(e) follow from the expansion of the expression of the tidal energy dissipation within the planet's rest frame in terms of the Runge-Lenz vector (see \citet{gu04} for more details) with $h_3(e) = (1 + 3e^2 + \frac{3}{8}e^4)(1-e^2)^{(-9/2)}$; $h_4(e) = (1 + \frac{15}{2}e^2 + \frac{45}{8}e^4 + \frac{5}{16}e^6)(1-e^2)^{(-6)}$ and  $h_5(e) = (1 + \frac{31}{2}e^2 + \frac{255}{8}e^4 + \frac{185}{16}e^6 +  \frac{25}{64}e^8 )(1-e^2)^{(-15/2)}$. Q$_p$ is the dimensionless tidal dissipation parameter of the planet \citep{gold66}; 
M$_{\star}$ is the mass of the star in solar units;  $R_p$ is the radius of the planet and a is the semi-major axis.\\

Using Eq.\ref{eq:eccxo3_ratio} we estimate the ratio of tidal heating dissipation rate to the insolation rate of XO-3b to be $ \frac{\dot{E}_{tide}}{\dot{E}_{insolation}}\vert _{e=\xothreeccwinn} \sim$ 0.43  given the \citet{winn08} planetary and stellar parameters: M$_{\star}$ = \xothreemsxot $\pm$ \xothreeemspxot M$_{\sun}$, M$_{p}$ = \xothreempxot $^{\xothreeemppxot}_{\xothreeempmxot}$ M$_{Jup}$, R$_{p}$ = \xothreerpxot $^{\xothreeerppxot}_{\xothreeerpmxot}$ R$_{Jup}$, $a$ = \xothreesemixot~$\pm$~\xothreeesemixot~AU, F$_p$ = 1.93 $\times 10^9$ erg cm$^{-2} s^{-1}$ , planetary tidal dissipation parameter Q$_{p}$ =10$^5$ and radial-velocity derived eccentricity of e = \xothreeccwinn $\pm$ \xothreeeccwinn. 

Our study refines the eccentricity of XO-3b e = \xothreecc $\pm$ \xothreeecc, which yields a ratio $ \frac{\dot{E}_{tide}}{\dot{E}_{insolation}}\vert _{e=\xothreecc} \sim$ 0.56 i.e. a 29\% increase in the tidal dissipation rate over the lower eccentricity when all other parameters are unchanged. Figure 2 of \citet{liu08} would suggest that if the age of XO-3b is currently estimated to be t = 2.82 $^{+0.58}_{-0.82}$ GYr \citep{winn08} and for solar metallicity, the increased tidal heating rate from our work would require a lowered tidal dissipation parameter Q$_p \lesssim$ 10$^6$.
Furthermore the radius-age trajectory for XO-3b with M$_{p}$ = \xothreempxot $^{\xothreeemppxot}_{\xothreeempmxot}$ M$_{Jup}$, R$_{p}$ = \xothreerpxot $^{\xothreeerppxot}_{\xothreeerpmxot}$ R$_{Jup}$ and the refined eccentricity e = \xothreecc $\pm$ \xothreeecc~is inconsistent with no tidal heating depicted by  infinite tidal heating parameter Q=$\infty$. 
 An important caveat to our radius interpretations for XO-3b is that we assume that tidal heating is the only radius inflation mechanism. Furthermore, the distance to XO-3b is still very uncertain (d = 260 $\pm$ 23 pc). Also, the discovery paper by \citet{xo3} and the photometric followup by \citet{winn08} disagree on the mass of XO-3b by 10\% and more than 50\% on the radius. Detailed radius-age determinations for XO-3b and its tidal heating history will thus need a parallax determination, which is in progress \citep{xo3_parallax}. Furthermore given the rudimentary nature of the Q model of tides, the assumption that dissipation is all in the convective core, and without detailed knowledge of the real physics of tidal dissipation, our conclusions about tidal heating rates and inferred radii of XO-3b are preliminary.\\ 

In short, we have refined the orbital eccentricity of XO-3b using the secondary eclipse timings in the 4 IRAC channels to e = \xothreecc $\pm$ \xothreeecc, which  increases the rate of tidal heating of the planet by 29\% over previous eccentricity estimates. Even in the absence of an accurate parallax measurement, the radius-age trajectory of XO-3b \citep[Fig.2 of][]{liu08} seems to imply that at least some amount tidal heating must be responsible for the inflated radius of XO-3b.

\subsection{Stratospheric temperature profile}
\label{anlxo3}

\begin{figure}
\centering
\includegraphics[totalheight=8cm]{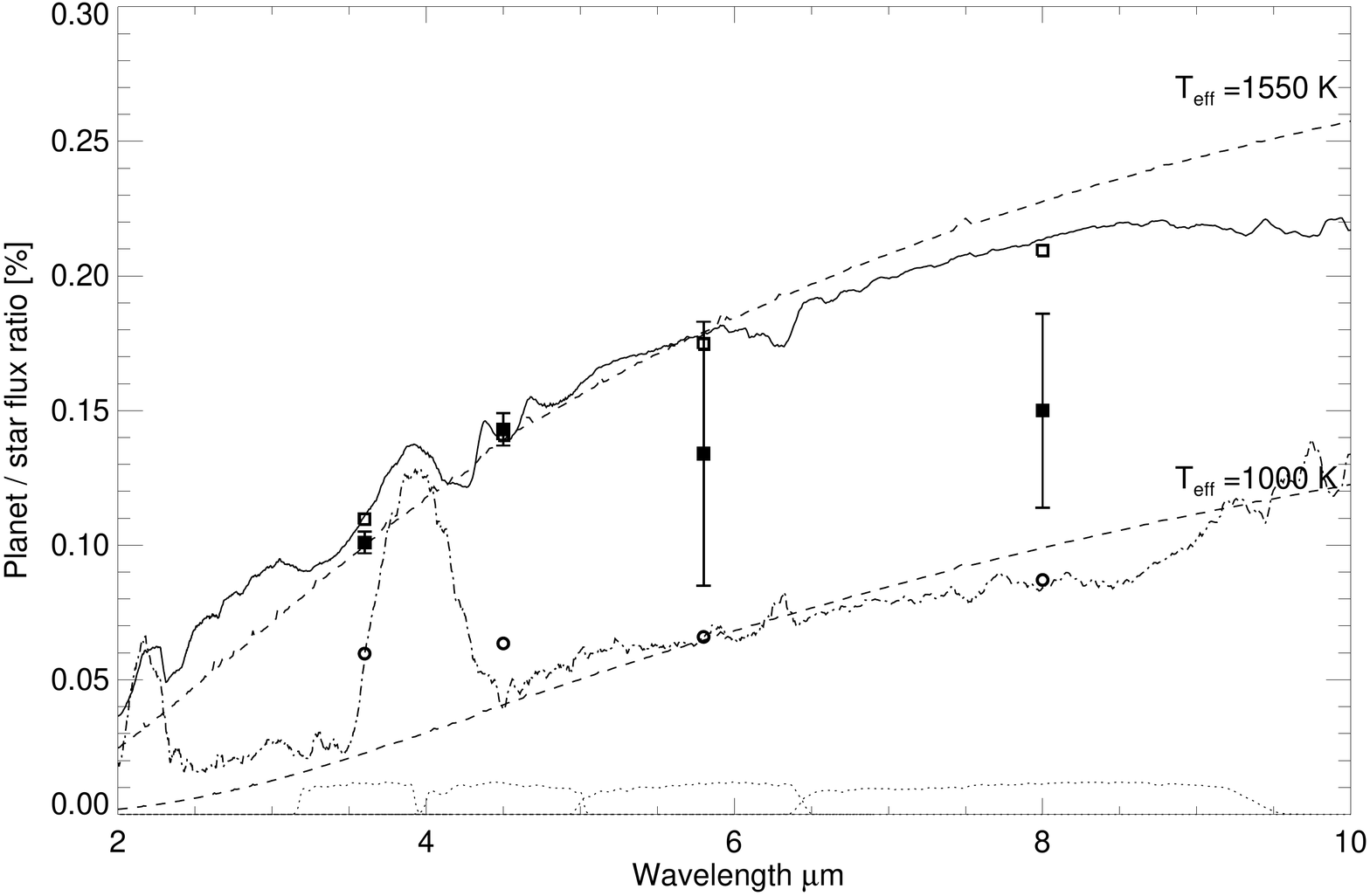}
\caption[XO-3b eclipse depths]{~\spit~IRAC secondary eclipse depths for \protect\objectname{XO-3b} with MCMC error bars (filled squares). The predicted emission spectrum of the planet \citep{burr07,burr07b,spiegel09} with an upper atmospheric absorber of $\kappa_{e}$ = 0.2 cm$^{2}$/g and redistribution parameter of P$_{n}$=[0.2] is plotted as a solid line. 
A model with no atmospheric absorber and a redistribution parameters of P$_{n}$=[0.3] is over plotted with dot-dashed line (see \S \ref{anlxo3} for details). Both model emission spectra are calculated for XO-3b orbital distance $a$ = \xothreesemixot~AU \citep{winn08} and stellar insolation $F_p \sim $\xothrinso $\times$ 10$^{9}$ ergs cm$^{-2}$ s$^{-1}$. See \S \ref{anlxo3} for more details. The band-averaged flux ratios are plotted as open squares and open circles for the models with and without an extra upper atmospheric absorber, respectively. The theoretical flux ratios obtained from a \protect\objectname{XO-3} stellar spectrum (from \protect\url{http://wwwuser.oat.ts.astro.it/castelli/grids/gridp05k2odfnew/fp05t6500g35k2odfnew.dat}) and an assumed black-body spectrum for the planet at [1000, 1550] K are plotted as dashed lines.
 The normalized~\spit~IRAC response curves  for the 3.6-, 4.5-, 5.8-, and 8.0~micron channels are plotted at the bottom of the figure (dotted lines).\label{fig:atmoxo3}}

\end{figure}

 The eclipse depths reported in Table \ref{tbl1xo3} and depicted as filled squares in Fig. \ref{fig:atmoxo3} show the spectral energy distribution of the upper atmosphere of XO-3b as a function of the flux ratio between XO-3b and its parent star XO-3. The flux ratio increases from the \one~to the \two~channels and stays constant within errors in the \three~and \four~channels. We compare the flux ratios (filled squares) to atmospheric models based on the methodology of \citet{burr07,burr07b,spiegel09}, which are depicted in Fig. \ref{fig:atmoxo3} as a black solid line (and open squares as IRAC band averages) and a dot-dashed line with open circles as IRAC band averages.  

The black solid line with open squares presents an atmospheric model with upper atmospheric temperature inversion induced by an extra absorber of uniform opacity of $\kappa_{e}$ = 0.2 cm$^{2}$/g placed at optical wavelengths 
and placed high up in altitude at pressures below P$_0$ = 30 mbars in XO-3b's atmosphere. 
The model incorporates a heat re-distribution parameter of P$_{n}$ = [0.2], which corresponds to an atmosphere between the two extremes of no heat re-distribution ( P$_{n}$=0) and full re-distribution  (P$_{n}$=0.5)  (see \citet{burr07b} for more details). 
The dot-dashed line with open circles as IRAC band averages corresponds to an atmospheric model with no extra upper atmospheric absorber and a heat re-distribution parameter of P$_{n}$ = [0.3]. \\
 Both atmospheric model are calculated for XO-3b's orbital distance $a$ = \xothreesemixot~AU \citep{winn08} and stellar insolation $F_p \sim $\xothrinso $\times$ 10$^{9}$ ergs cm$^{-2}$ s$^{-1}$, which ignores the dynamical atmospheric effects due to variable stellar insolation caused by XO-3b orbital eccentricity $e = \xothreecc \pm \xothreeecc$.
A full dynamical model for the atmosphere of XO-3b, which incorporates the time variable stellar insolation and the temporal adjusteent of the atmosphere to the instanteneous irradiation (i.e. the radiative time constant), is beyond the scope of this paper. Full dynamical treatment of XO-3b's atmosphere is planned for the dynamic weather observations of XO-3b during the Warm Spitzer mission in the spring of 2010 in the \one~and~\two~IRAC channels. 

The planet/star contrast ratios of XO-3b are within [\onesig,\twosig, \threesig, \foursig]-$\sigma$ of the thermal inversion in the upper stratosphere model of XO-3b in the \one~,\two,~\three,~and \four~channels respectively.
~The measured planet/star contrast ratios are inconsistent at more than 3-$\sigma$ in the \one, \two, \three,~and \four~channels with the thermally non-inverted upper atmosphere model (dot-dashed line and open circles as band averages in Fig. \ref{fig:atmoxo3}). The flux contrast ratios of XO-3b in the 4 IRAC channels thus represent a detection of an upper atmospheric temperature inversion similar to the temperature-pressure profile depicted as a solid line in Fig. \ref{xo3b_tp}. 
We further note that the XO-3b flux ratios can be reproduced in the \one, \two,~and \three~channels by a black body with an effective temeprature T$_{eff}$ = 1550 K as well. \\

A correlation between minimum insolation at the planet's substellar
 point and the presence of stratospheric temperature inversions has been
 recently emerging \citep{burr07b,fort07b} from numerous Hot-Jupiter
 spectral energy distribution measurements \citep{har07,charb08,knutson07b,knutson_tres4,machalek08,machalek09,odonovan09,fressin09,todorov09,christi09,gill09}. 
Currently $F_p \sim$ 1.0$\times$ 10$^{9}$ ergs cm$^{-2}$ s$^{-1}$ of flux at the planetary substellar point is thought to be be necessary for the extra optical absorber to drive a stratospheric temperature  inversion, although significant outliers exist: XO-1b with a substellar point flux of $F_p \sim$ 0.49$\times$ 10$^{9}$ ergs cm$^{-2}$ s$^{-1}$ has a stratospheric temperature inversion \citep{machalek08}; while TrES-3 is strongly irradiated and yet posseses no thermal inversion according to \citet{fressin09}. 
The planet HAT-P-1b has intermediate subsolar flux between XO-1b and TrES-3 and presents evidence for a weak thermal inversion \citet{todorov09}. Also, the flux ratios of the planet CoRot-2b ($F_p \sim $1.3 $\times$ 10$^{9}$ ergs cm$^{-2}$ s$^{-1}$) in \two~and \four~IRAC channels provide a tentative non-detection of thermal inversion  \citep{gill09}.

The distance to XO-3 is currently uncertain, so are the estimates for the stellar mass and radius. Therefore the substellar point flux at the XO-3b is estimated to be in the range $F_p \sim $(1.9 - 4.2) $\times$ 10$^{9}$ ergs cm$^{-2}$ s$^{-1}$, \citet{liu08}. 
This entire flux range is well above the threshold value and therefore strongly
 predictive of a temperature inversion in the stratosphere of XO-3b,
 which is detected in our dataset.\\

The diagnosis of temperature inversions in Hot Jupiter atmospheres is still somewhat model dependent as exemplified by the color-color diagram of \citet{gill09}.
 This figure shows that although TrES-3b \citep{fressin09} and TrES-2b \citep{odonovan09} have almost identical colors, an inversion is claimed for TrES-2b but not for TrES-3b.  
Alternative determinants for the cause of temperature inversions in stratospheres of hot Jupiters have been suggested by \citet{zahn09} in the form of sulfur photochemistry. Furthermore, three dimensional global circulation models (3D GCM) by \citet{showman09,rauscher09} suggest that dynamic weather patterns can induce temperature inversions even without extra stratospheric optical absorbers. Obtaining flux ratios of hot Jupiters with varying degrees of stellar insolation, planetary metallicity, and eccentricity at multiple IR wavelengths with Spitzer IRAC or JWST in the future will help to constrain the cause of stratospheric thermal inversions in hot Jupiters.  

\section{Conclusion}
We determined the flux ratios of the planet heavy and eccentric planet XO-3b to its parent star in the 4 IRAC bands: \xothreeflone~$\pm$ \xothreeeflone~at \one; \xothreefltwo~$\pm$ \xothreeefltwo~at \two; \xothreeflthr~$\pm$ \xothreeeflthr~at \three~and \xothreeflfour~$\pm$ \xothreeeflfour~at \four. The flux ratios point towards a stratospheric temperature inversion best fit with atmospheric models with a uniform stratospheric absorber of $\kappa_{e}$ = 0.2 cm$^{2}$/g. \\

XO-3b is strongly irradiated with a subsolar point flux $F_p \sim $(1.9 - 4.2) $\times$ 10$^{9}$ ergs cm$^{-2}$ s$^{-1}$, depending on uncertain parent star parameters and eccentric orbit. This high flux is expected to cause a thermal inversion in the planet's stratosphere, which is indeed observed. 
Obtaining the parallax distance \citet{xo3_parallax} to the parent star XO-3 would refine both the stellar and planetary masses and radii and hence constrain better the subsolar point flux $F_p$. 
The correlation between the presence of a temperature inversion in a hot Jupiter atmosphere and the subsolar point flux from the parent star is insufficient to explain why high insolation planets like TrES-3 do not have stratospheric inversions and some low insolation planets like XO-1b do have inversions. Secondary factors such as sulfur chemistry, atmospheric metallicity, amounts of macroscopic mixing in the stratosphere, or even dynamical weather effects likely play a role.\\

Using the secondary eclipse timing centroids we refined the orbital eccentricity of XO-3b to be  e = $\xothreecc \pm \xothreeecc$, which is 1.0-$\sigma$ higher than the radial velocity derived eccentricity e = \xothreeccwinn $\pm$ \xothreeeccwinn \citep{winn08,xo3}. The refined eccentricity increases the amount of tidal energy dissipation rate by 29\%, and the radius-age trajectories for XO-3b thus imply that at least some amount of tidal-heating must be responsible for the inflated radius of XO-3b. The tidal heating parameter is constrained to Q$_p \lesssim$ 10$^6$. 
A more accurate radius measurement of XO-3b is needed from a parallax
 distance to the parent star XO-3 either from the Hubble Space Telescope
 or the future GAIA mission to further refine its tidal heating rate
 and the allowable range for the tidal heating parameter Q$_p$. 

\clearpage

\begin{deluxetable}{ccccc}
\tablecolumns{5}
\tablewidth{0pt}
\tablecaption{Secondary eclipse best fit parameters for XO-3b}
\tablehead{
\colhead{$\lambda$}  &\colhead{Eclipse Depth $\Delta F$} &\colhead{Eclipse Center Time} &\colhead{Eclipse Center Phase  } \\
\colhead{(microns)}  &\colhead{} &\colhead{(HJD)} &\colhead{$\Phi$}}
\startdata
3.6  &\xothreeflone~$\pm$ \xothreeeflone & \xothreehjdone~$\pm$ \xothreeehjdone& \xothreephaseone~$\pm$ \xothreeephaseone\\
4.5  & \xothreefltwo~$\pm$ \xothreeefltwo & \xothreehjdtwo~$\pm$ \xothreeehjdtwo& \xothreephasetwo~$\pm$ \xothreeephasetwo\\
5.8  & \xothreeflthr~$\pm$ \xothreeeflthr & \xothreehjdthr~$\pm$ \xothreeehjdthr& \xothreephasethr~$\pm$ \xothreeephasethr\\
8.0  & \xothreeflfour~$\pm$ \xothreeeflfour & \xothreehjdfour~$\pm$ \xothreeehjdfour& \xothreephasefour~$\pm$ \xothreeephasefour\\
\enddata
\label{tbl1xo3}
\end{deluxetable}

\acknowledgements
The authors would like to thank the annonymous referee for a speedy and thorough review, which has substantially improved the manuscript. The authors would also like to acknowledge the use of publicly available routines by Eric Agol and Levenberg-Marquardt least-squares minimization routine MPFITFUN  by Craig Markwardt.  P.M. and P.R.M. were supported by the Spitzer Science Center Grant C4030 to the Space Telescope Science Institute and the Bay Area Environmental Research Institute.
A.B. was supported in part by
NASA grant NNX07AG80G.  We also acknowledge support through
JPL/Spitzer Agreements 1328092, 1348668, and 1312647. T.G. acknowledges funding by NASA Ames Research Center to the Ames
 Center for Exoplanet Studies in support of this work.
 This work is based on observations made with the Spitzer Space Telescope, which is operated by the Jet Propulsion Laboratory, California Institute of Technology under a contract with NASA. This publication also makes use of data products from the Two Micron All Sky Survey, which is a joint project of the University of Massachusetts and the Infrared Processing and Analysis Center/California Institute of Technology, funded by the National Aeronautics and Space Administration and the National Science Foundation. \\

\end{document}